# Engineering Microbial Symbiosis for Mars Habitability

**RANDALL R. CORRELL[1], SIMON P. WORDEN[1]**
[1]Zetator UG, Saarstr. 26, 54290 Trier, Germany

**Email** correll@zetator.com

The colonization of Mars presents extraordinary challenges, including radiation exposure, low atmospheric pressure, and toxic regolith. Recent advancements in synthetic biology and genetic engineering offer unprecedented opportunities to address these obstacles by utilizing terrestrial extremophiles and engineered organisms. This paper examines the potential for creating symbiotic relationships between terrestrial microbes and hypothetical Martian life forms, should they exist, to support a sustainable human presence on Mars. Inspired by natural examples of endosymbiosis, such as mitochondria and chloroplasts, we propose methods to engineer life forms capable of enduring Martian conditions. Key components include experimental designs, laboratory simulations, and bioengineering approaches essential to this endeavor. The ethical, political, and technological challenges of introducing engineered life to Mars are critically evaluated, with an emphasis on international collaboration and robust planetary protection policies. This research underscores engineered symbiosis as a transformative strategy for enabling life to adapt and thrive on Mars while advancing humanity's aspirations for interplanetary habitation and exploration. By addressing these challenges, this work highlights a path toward sustainable life on Mars, reflecting both scientific ingenuity and ethical stewardship.

## 1. Introduction

Mars has long captured human imagination and scientific curiosity. Once thought to harbor alien civilizations, as depicted in Percival Lowell's speculation about Martian canals [1], it remains a focal point for planetary exploration. While early fantasies have faded, modern missions have revealed an apparently barren yet scientifically promising planet—a world with the potential to deepen our understanding of life and its adaptability.

Serious exploration began with Mariner 4's flyby in 1965, exposing a cratered, desolate surface. NASA's Viking lander in the 1970s was perhaps the first to encounter, unknowingly, perchlorates in the Martian soil [2]. Then NASA's Phoenix lander in 2008 advanced our knowledge, identifying perchlorates in Martian soil [3]. This is a discovery signaling both the challenges and opportunities of habitability, as perchlorates are detrimental to many organisms on the Earth, and the origins of perchlorates on Mars are not fully understood [4]. Subsequent missions, including the Curiosity and Perseverance rovers, have provided evidence of ancient water, organic molecules, and seasonal methane variations, reinforcing the possibility of past or present microbial life [5][6]. And in recent decades, proponents have become more active in developing plans for settlements on Mars [7].

Yet Mars remains a hostile environment. Its thin, carbon dioxide-dominated atmosphere, high radiation levels, and toxic regolith create formidable barriers to human survival and colonization [8][9]. To address these challenges, researchers have turned to bioengineering to enable and accelerate the habitability of Mars. These approaches include biological consortia [10], partial gravity experiments of the Eu:CROPIS mission [11], synthetic biology for more efficient photosynthesis [12], approaches to microbial manufacturing that also explores the problem of perchlorates on Mars[13], and especially the use of extremophiles—organisms thriving in Earth's most inhospitable environments—for the dissolution of carbonate rocks [14]. By integrating natural adaptations from extremophiles with the new techniques of synthetic biology, researchers aim to design organisms capable of enduring Mars' severe conditions.

This paper examines the feasibility of engineering Mars-adapted organisms for habitability, proposing experimental frameworks, laboratory models, and bioengineering strategies. Ethical and political implications of introducing engineered life to Mars are addressed, emphasizing planetary protection and international collaboration. By combining extremophile resilience with synthetic biology precision, engineered symbiosis offers a bold approach to overcoming Mars' challenges and advancing humanity's interplanetary aspirations. This work is becoming more pressing as government and commercial efforts accelerate plans to visit Mars [15][16], and techniques to enable and accelerate *in situ* resource utilization [17][18] will be needed.

In 1999, the co-authors of this paper visited Sri Lanka, where they had the privilege of meeting Arthur C. Clarke. Over an extraordinary week, they delved into profound discussions on the limits of life, the concept of terraforming Mars, and the nature of extraterrestrial organisms. Clarke's insights profoundly influenced their thinking. Later meetings, including his 2002 visit to the United States, solidified their shared vision for humanity's future beyond Earth. Leveraging modern AI, the authors have developed a digital avatar of Clarke, an AI assistant, named ArthurGPT, grounded in his writings and interviews. This avatar contributed to the drafting of this paper, reflecting Clarke's visionary perspective on Mars exploration. The Appendix explains how the human authors collaborated with ArthurGPT in developing this paper.

# 2. Background

## 2.1 Adapting Terrestrial Extremophiles for Life on Mars

*Mars-Adapted Organisms via Endosymbiosis*

The endosymbiotic theory, first proposed by Lynn Margulis in the 1960s [19], describes how certain organelles—such as mitochondria and chloroplasts—originated as free-living bacteria that were engulfed by ancestral eukaryotic cells and formed a stable, mutually beneficial relationship. A figure of a typical eukaryotic cell with cellular components such as nuclear DNA, RNA, mitochondria, and ribosomes is shown in Figure 1 [20]. If one considers engineering extremophiles for Mars, extremophiles from the archaea domain of life (archaea, bacteria, eukarya [21]) are excellent candidates due to their resilience, but their simplicity might limit their functionality for complex bioengineered habitats. However, a hybrid approach, incorporating extremophiles into eukaryotic cells via endosymbiosis (or using their genes in engineered eukaryotic systems), could be an interesting strategy.

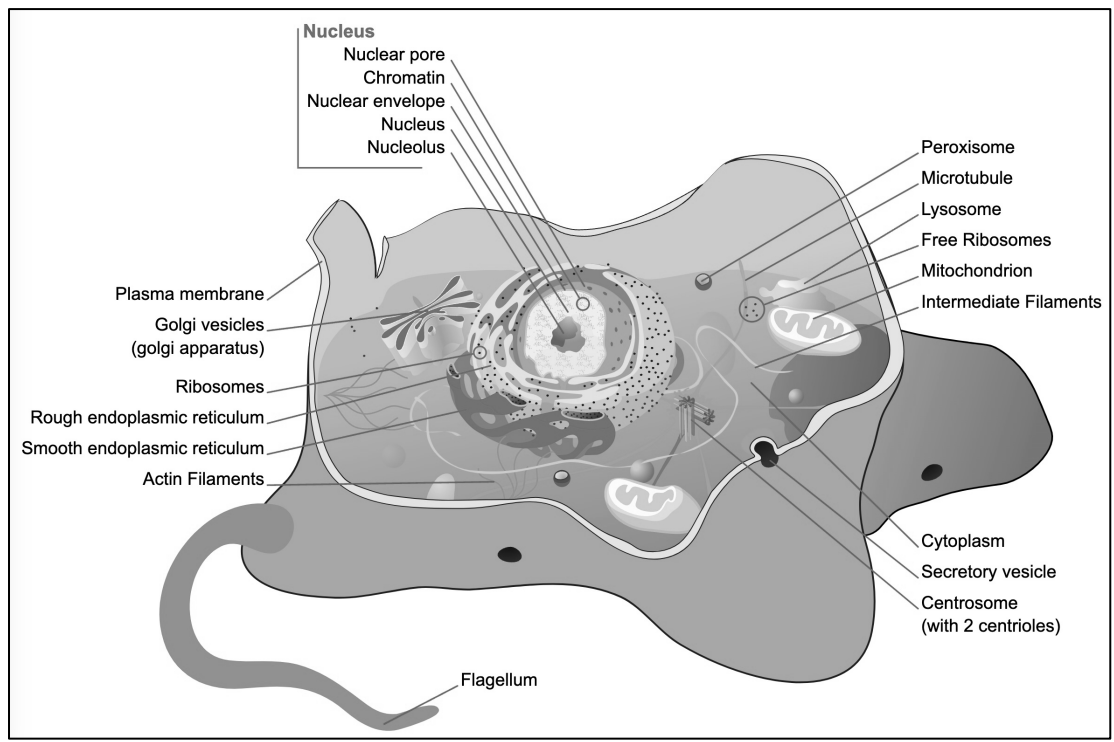

*Figure 1. Eukaryote cell with various cellular elements depicted* [20]. *The figure shows the approximate size and scale of the cell's components and organelles, including the mitochondria, thought to be an endosymbiont from the archaea domain of life.*

*Extremophiles*

Earth's extremophiles offer remarkable insights into life's adaptability, serving as critical models for engineering organisms suited to Mars' severe conditions. Extremophiles thrive in environments extreme by Earthly standards: high radiation, temperature extremes, acidic or alkaline pH, salinity, and anaerobic conditions [22][23]. These organisms showcase biochemical pathways and genetic adaptations vital for survival under immense stress, providing a blueprint for creating life forms capable of enduring extraterrestrial environments.

*Notable Examples of Extremophiles*

Deinococcus radiodurans: Dubbed "Conan the Bacterium," Deinococcus radiodurans tolerates radiation thousands of times more intense than what would be lethal to humans. [24] This resilience stems from extraordinary DNA repair mechanisms, allowing it to reassemble its genome hours after radiation damage. Proteins such as RecA and PprA safeguard its DNA and proteins from oxidative harm, facilitating double-strand break repair [25]. These mechanisms inspire approaches to enhance radiation resistance in other organisms, a critical adaptation for Mars' high-radiation environment.

Halobacterium salinarum: Thriving in hypersaline habitats, Halobacterium salinarum survives where few organisms can. Specialized proteins, such as halorhodopsin (a light-driven ion pump) and bacteriorhodopsin (which converts light into chemical energy), maintain osmotic balance. These adaptations enable survival under desiccation and salinity, conditions akin to Martian regolith. Additionally, its carotenoid pigments shield it from UV radiation, offering potential for engineering Mars-tolerant organisms [26].

Pyrococcus furiosus: Found in hydrothermal vents, Pyrococcus furiosus endures temperatures above 100°C. Its extremozymes remain stable and functional under extreme heat, useful in biofuel production and biocatalysis. Enzymes like its DNA polymerase suggest strategies for designing proteins that function in Martian temperature fluctuations [27]. Its thermophilic metabolism illustrates pathways for engineering resilience to harsh conditions.

*Practical Applications of Extremophiles*

Extremophiles' significance extends beyond theory to practical biotechnology. Enzymes from extremophiles—like Thermus aquaticus DNA polymerase used in PCR—have revolutionized molecular biology. Extremozymes from Pyrococcus furiosus and Halobacterium salinarum support industrial processes under high heat, pressure, or salinity. And ideas have recently been proposed to use microbial mats of extremophiles on Mars [28].

*Endosymbiosis of Earth Organisms to Thrive on Mars*

For Mars colonization there are several key extremophile abilities that one would want to integrate into the new habitable Martian biome. Radiation resistance mechanisms from D. radiodurans [24] could be integrated into plants or microbes to counteract Mars' UV radiation. Salt tolerance pathways from H. salinarum may help organisms metabolize Mars' perchlorate-rich soils [29]. And heat-stable enzymes from P. furiosus could support biochemical functions in fluctuating thermal conditions [30].

Through extremophile adaptations, scientists can design hybrid organisms capable of surviving Mars' environment. These top-level strategies would be implemented using a variety of genetic engineering and synthetic biology toolkits.

## 2.2 Exploiting Martian Life

### *Panspermia*

If there happens to be life on Mars, there are two possibilities of how to collaborate with it. The first possibility is that Martian organisms use genetic codes like terrestrial DNA and RNA. This would be a happy situation; in that it is known how they work. And it might be more than a coincidence. It has been proposed that all life on earth originated from noncellular entities dubbed First Universal Common Ancestor (FUCA) thought to use RNA for a genetic code and self-replication. Such a process may also have occurred on Mars.

It is also likely that the three terrestrial planets, Venus, Earth and Mars exchanged substantial material, particularly in their early first billion years [31]. This concept is known as panspermia. If this material contained earth life, LUCA or FUCA, it's possible that hypothetical Mars (or Venus) life could be DNA or RNA based.

Some experts suggest that life, or at least basic biochemistry, has an interstellar origin, perhaps in interstellar clouds [32]. In this case life on the Earth and Mars would be expected to have common chemical origins and more likely to be compatible.

### *DNA/RNA-Based Genetic Information*

DNA in membrane-bound organelles, such as mitochondria and chloroplasts, does not mix with host-cell DNA. Mars organisms may or may not have membrane-bound organelles, but the Martian cell itself would presumably have a membrane, and just as early Eukarya was able to form a symbiosis, such a symbiosis might be engineered where the entire Mars cell becomes a new, implanted organelle. Conversely Mars life may not have organelles but may include non-membrane-bound organelles such as RNA-powered ribosomes with unique, desirable properties. The genetic code of these might be extracted from Martian life and inserted into earth-cell nuclear DNA to enable these properties to be included in the hybrid cell.

### *Alien Genetic Information*

The second possibility, if life is found on Mars, is that it is non-DNA/RNA based. In his seminal 1944 book, "What is Life?", Erwin Schoedinger introduced the idea of an aperiodic solid that contained genetic information in its configuration of covalent chemical bonds [33].

It is possible that alien life might have fundamentally different nucleic chemistry. Seagar suggest that hypothetical Venus's cloud life, which would have to survive in high sulfuric acid conditions, possibly using the acid itself as a solvent, could use a genetic material composed of a DNA backbone with amino acid pairs, rungs on the DNA ladder different than the four amino acids in terrestrial nucleic acids but stable in the high acid Venusian environment [34]. Such a genetic code approach may have had common origins as terrestrial life but over time and changing conditions mutated to a new form more suitable to the new conditions. Such a process may have occurred on Mars as well. If the alien conditions are contained within a membrane-bound unit, it might also be suitable for implantation or could be modified for use within earth organisms. Key is to locate and sequence hypothetical Mars life.

Whatever life might be discovered on Mars, it clearly would accelerate the ability to create enhanced forms of Earth and Martian hybrid organisms to support habitability.

## 2.3 Genetic Engineering Techniques

Advances in genetic engineering revolutionize biology, enabling precise organism modifications. Techniques like CRISPR-Cas9, synthetic biology, and horizontal gene transfer open new possibilities for engineering life forms adapted to Mars.

### *CRISPR-Cas9 and Genome Editing*

CRISPR-Cas9, derived from bacterial defense systems, enables precise genome editing by targeting specific DNA sequences. It has been used to introduce stress-resistance traits in plants, such as drought tolerance [35]. CRISPR-Cas9 could enable microbes and plants to withstand the high levels of ionizing radiation present on the Martian surface. Additionally, engineered microbes could be designed to metabolize perchlorates as an energy source, turning a toxic compound into a useful resource for sustaining microbial life. Furthermore, CRISPR could facilitate the development of nitrogen-fixing bacteria tailored for Martian agriculture, enriching the regolith to support plant growth in extraterrestrial environments.

### *Horizontal Gene Transfer*

Horizontal gene transfer (HGT) allows genetic material movement between organisms without reproduction. Scientists harness HGT to transfer traits like salinity tolerance or UV resistance. Genes from extremophiles such as Halobacterium salinarum could be transplanted to terrestrial organisms, creating hybrids adapted to Mars.

### *Organelle Engineering*

Life began on earth around 4 billion years ago—within a few hundred million years of the earth cooling enough for life to exist. The earliest life form, from which all present terrestrial life has evolved, is known as the Last Universal Common Ancestor or LUCA. It evolved into two branches, Archaea and Bacteria. Archaea differ from Bacteria in several key ways, including cell membrane chemistry and more diverse metabolic pathways. They were not identified as separate from Bacteria unto 1977 [21]. Archaea's diverse metabolic pathways enable them to thrive in more diverse environments. Most extremophiles, including those than can metabolize perchlorates, are archaea. Interestingly, LUCA contained cell powerplants for protein synthesis called ribosomes, as all branches of life contain them. These ribosomes are the earliest form of organelles.

The third branch of life is Eukarya. Between 1-2 billion (or more) years ago a symbiosis occurred where one or more Archaea consumed but did not destroy Bacteria [36]. The combined cell continued to evolve so that the bacteria formed various functions with the cell, mostly energy production [37]. These became organelles, each with its own, separate genetic material, which also evolved. Eukarya could contain thousands of these organelles, in plants these are the chloroplasts able to convert sunlight into chemical energy, and in other cells, including animal cells, these became energy producing mitochondria. The great increase in available energy enables greatly increased cell complexity, including multicellular organisms.

A milestone in genetic engineering involves transplanting organelles across species. Kuroiwa demonstrated chloroplast integration into mammalian cells, enabling photosynthesis [38]. This suggests the potential for incorporating hypothetical Martian microbial organelles with radiation or perchlorate metabolism into terrestrial cells, creating hybrid organisms for Mars. The latter would represent engineered solutions much better adapted for the Martian environment.

A key factor in this will be finding useful biological approaches that also survivable in the biome and evolutionarily stable. While the example above of enabling photosynthesis in mammalian cells was successful, it was not long lasting, providing its new functionality for only a few days [38][39]. A key objective of this paper is to present how such efficacious and stable adaptations might be accomplished.

### *Synthetic Biology and Biodesign*

Synthetic biology enables the design and construction of entirely new biological systems tailored for specific tasks. This includes bacteria engineered to synthesize biofuels and bioplastics, plants modified with marine genes to exhibit bioluminescence, and microbial consortia designed to perform complex biochemical tasks such as nitrogen fixation.

On Mars, synthetic biology could play a critical role in establishing a sustainable human presence by enabling biological systems to perform essential functions. Engineered microbes could metabolize perchlorates in the Martian soil while simultaneously releasing oxygen, mitigating toxicity and contributing to atmospheric enrichment. Plants could be designed to fix nitrogen in the regolith, enhancing its fertility for future agricultural use. Additionally, carefully constructed microbial ecosystems could support human habitation by recycling waste, producing essential biomolecules, and maintaining environmental stability.

## 2.4 Previous Mars Missions

Mars exploration has unveiled critical data on its habitability and potential to support life. Over five decades, robotic missions have provided insights into Mars' geology, atmosphere, and chemistry.

### *Key Mars Missions*

There have been a number of important Mars missions that have provided invaluable data to understanding the environment on Mars.

- Viking Program (1976): Conducted life detection experiments and identified perchlorates in Martian soil, highlighting challenges and opportunities for microbial survival.
- Mars Odyssey (2001): Mapped subsurface hydrogen, indicating water ice.
- Curiosity Rover (2012-present): Discovered ancient streambeds, organic molecules, and methane, suggesting Mars once had habitable conditions.
- Perseverance Rover (2021-present): Collecting rock and soil samples for return to Earth and searching for ancient microbial life in Jezero Crater.
- ExoMars Mission (2022): Drilled beneath Mars' surface for biosignatures protected from radiation.

### *Contributions to Understanding Mars' Habitability*

These missions confirmed Mars once had liquid water and potentially habitable conditions while identifying key challenges: radiation, low atmospheric pressure, and toxic soil chemistry. The discovery of perchlorates underscores the need for bioengineering strategies to mitigate toxicity and harness energy potential.

By integrating findings from extremophiles, genetic engineering, and synthetic biology, researchers can address Mars' challenges, advancing sustainable colonization strategies.

## 3. Scientific Methods and Models

The development of engineered symbiosis between Martian and terrestrial organisms demands a methodical, interdisciplinary approach. This section outlines experimental designs, laboratory simulations, and bioengineering strategies essential for creating hybrid organisms capable of enduring Mars' extreme conditions. By integrating tools from genetics, synthetic biology, and environmental engineering, researchers can establish models to test and optimize these life forms.

## 3.1 Experimental Designs

Creating engineered symbiosis requires rigorous experimental processes to identify, analyze, and test potential hybrid organisms. Hypothesis-driven research and empirical testing must align with biological and environmental constraints.

### *Step 1: Culturing Hypothetical Martian Microbes*

If Martian microbes are discovered, the first step involves their isolation and cultivation under controlled laboratory conditions. Specialized equipment is required to accommodate potentially unique biochemical and physiological properties. Developing biological systems for Mars requires precise control over multiple factors, including growth medium composition, environmental conditions, and containment measures. Growth medium design must optimize nutrient availability, potentially utilizing Martian regolith analogs that contain perchlorates and essential minerals to support microbial and plant growth. Environmental control is crucial, necessitating simulations of Martian temperature fluctuations, atmospheric pressure, radiation exposure, and humidity levels to ensure biological viability under extraterrestrial conditions. Additionally, stringent sterility and containment protocols must be established to prevent unintended contamination of Earth's biosphere, aligning with planetary protection policies and ensuring responsible exploration and habitation strategies.

In the absence of confirmed Martian life, terrestrial extremophiles like Deinococcus radiodurans and Halobacterium salinarum provide useful analogs. Insights from extremophiles help shape strategies for managing and studying hypothetical Martian organisms.

### *Step 2: Genome Sequencing and Bioinformatics*

Cultured Martian microbes must undergo genome sequencing to identify survival traits. Advances in high-throughput sequencing technologies, such as Illumina and PacBio systems, enable efficient analysis of microbial genomes. The objectives of this research focus on identifying and characterizing genetic traits essential for survival in Martian conditions. Gene identification efforts will target key adaptations such as radiation resistance, perchlorate metabolism, and oxidative stress tolerance. Comparative genomics will be employed to analyze potential Martian genomes in relation to terrestrial extremophiles, identifying both conserved and novel genetic pathways relevant to extraterrestrial habitability. Additionally, functional annotation using bioinformatics tools such as BLAST and KEGG will facilitate the prediction of gene functions and metabolic networks, providing insights into the biochemical potential of Martian and engineered microbial systems.

These findings will guide genetic engineering efforts to transfer Martian survival traits into terrestrial organisms, forming bioengineered hybrids.

*Step 3: AI and Quantum Computational Design*

Advanced computing techniques such as AI and quantum computing are increasingly used in synthetic biological design to alter gene expression, find alternative metabolic pathways, and even add new sequences to genetic code. Advanced deep learning models, including GPTs and reinforcement learning algorithms, are increasingly utilized in synthetic biology to design novel genetic sequences. For instance, the Evo model, trained on 2.7 million microbial and viral genomes, functions as a biological Rosetta Stone, enabling the design of synthetic CRISPR gene-editing tools and DNA sequences exceeding one million base pairs [40]. Similarly, GenerRNA, a Transformer-based model, facilitates de novo RNA design without requiring prior structural knowledge, broadening applications in vaccines and therapeutics [41]. These advancements underscore the transformative role of deep learning in engineering genetic code. Similarly, quantum computing algorithms have been developed to better compare genetic sequences [42]. Together these new advanced computing capabilities for machine learning and quantum computing will accelerate the design and implementation of new synthetic biological organisms.

*Step 4: Simulation Studies*

Simulated Martian conditions are essential for evaluating the viability of hybrid organisms, offering critical insights into survival rates, physiological thresholds, and metabolic adaptations. Atmospheric simulation replicates Mars' $CO_2$-dominant atmosphere and its extremely low pressure, approximately 0.6% of Earth's. Temperature extremes must be modeled to reflect the planet's diurnal fluctuations, which range from −125°C at night to 20°C in daytime. To assess radiation tolerance, controlled exposure to gamma radiation sources can simulate the high levels of ionizing radiation present on Mars' surface. Chemical composition studies integrate Martian regolith analogs containing perchlorates and sulfur compounds into growth media, enabling researchers to examine microbial and plant interactions with the planet's geochemical environment. These controlled experiments provide a foundation for designing robust biological systems capable of functioning under extraterrestrial conditions. Additionally, studies could test microbial metabolism of perchlorates as an energy source, informed by Earth-based perchlorate-reducing bacteria [43].

## 3.2 Laboratory Simulations

Replicating Martian conditions in laboratory simulations is essential for effective bioengineering tests. Several global facilities simulate key Martian environmental factors, offering controlled platforms for experimentation, although it must be noted that some Martian soil simulants do not contain perchlorates for safety reasons, and the perchlorates would need to be safely added to the simulant before laboratory experiments could be carried out..

*Key Facilities*

- Mars Simulation Laboratory (Denmark): Replicates Martian atmospheric pressure, temperature, and regolith composition for studying microbial and plant survival. [44]
- NASA Ames Research Center (USA): Provides Mars analog facilities to simulate Martian soil and environmental stresses, including radiation exposure. [45]
- Yungay Valley, Atacama Desert (Chile): As one of Earth's driest regions, it serves as a Mars analog for microbial survival and biogeochemical field studies. [46]

*Research Focus Areas*

Laboratory simulations facilitate testing bioengineered organisms in areas such as:

- Metabolic Adaptation to Perchlorates: Exploring how engineered microbes utilize perchlorates as an energy source, similar to Dechloromonas species.
- Radiation Resistance Mechanisms: Assessing integration of radiation-resistant traits from extremophiles like Deinococcus radiodurans.
- Symbiotic Relationships: Evaluating interactions between terrestrial and Martian microbes for mutual benefits.
- Plant-Microbe Interactions: Testing microbial consortia's ability to enhance plant growth in Martian regolith for agriculture.

By combining terrestrial Mars analogs with controlled experiments, researchers refine survival models for Martian environments.

## 3.3 Bioengineering Approaches

Creating hybrid organisms suited for Mars relies on cutting-edge bioengineering strategies. These approaches, inspired by natural processes like horizontal gene transfer and endosymbiosis, leverage modern synthetic biology tools.

*Horizontal Gene Transfer*

Horizontal gene transfer (HGT) introduces desirable traits across species. For instance, genes from Deinococcus radiodurans that confer radiation resistance could be inserted into Escherichia coli. Similarly, perchlorate-reducing pathways could be integrated into terrestrial microbes, enabling metabolism of Martian soil components.

*Organelle Integration*

Inspired by endosymbiosis, organelle engineering enables cross-species integration of organelles. Recent successes, such as chloroplast transplantation into mammalian cells [38], suggest potential applications. Hypothetical Martian organelles could be integrated into terrestrial cells to enhance their ability to survive in Mars' extreme environment. These engineered organelles could enable cells to utilize perchlorates as an alternative energy source, leveraging Martian soil chemistry to support metabolic processes. Additionally, they could provide resistance to high oxidative stress and elevated radiation levels, mimicking the adaptive mechanisms of extremophiles.

*Synthetic Consortia*

Designing microbial consortia involves engineering cooperative ecosystems with complementary functions. Nitrogen fixation could be facilitated by genetically modified microbes capable of enriching Martian soil, making it viable for agricultural use. Photosynthetic microorganisms could be deployed to convert atmospheric $CO_2$ into oxygen, supporting both human respiration and fuel production. Additionally, regolith bioremediation strategies would focus on microbial detoxification of perchlorate-laden Martian soil, improving its suitability for plant growth. By constructing robust microbial systems, researchers can develop self-sustaining ecosystems to support human colonization and resource utilization on Mars.

## 4. Ethical and Political Considerations

The prospect of colonizing Mars and bioengineering life for its harsh environment raises profound ethical and political questions. These discussions are essential to align exploration and colonization with moral, environmental, and geopolitical values. This section examines the ethical debates, planetary protection policies, and international cooperation necessary for Mars colonization.

### 4.1 Ethical Debates

Mars colonization presents complex ethical implications, encompassing humanity's responsibility as stewards of other worlds, the potential for disrupting alien ecosystems, and the philosophical challenges of interplanetary expansion.

*Is it Ethical to Prioritize Human Colonization Over Preserving Alien Life?*

A central dilemma involves balancing human expansion with the preservation of extraterrestrial life, if it exists. Advocates argue that humanity has a moral imperative to expand beyond Earth to ensure long-term survival. Threats like climate change, resource depletion, and catastrophic events highlight Mars as a viable backup for human civilization [47]. Supporters also emphasize benefits such as technological innovation, economic growth, and the inspiration derived from exploration.

Critics caution that colonization could irreversibly harm Martian ecosystems, especially if microbial life exists. Even dormant or minimal life forms raise ethical concerns about disrupting Mars' natural state. Discovering native Martian life would represent a profound scientific and philosophical milestone, potentially outweighing colonization interests [48].

Additionally, critics question the motivations behind colonization. Does Mars colonization reflect a scientific mission, or is it a colonialist endeavor prioritizing human interests over planetary integrity? Ethical approaches must learn from Earth's history of ecological exploitation, approaching Martian environments with humility and respect.

*Safeguards to Prevent Ecological Harm*

Robust safeguards are essential to address ecological concerns and preserve Mars' natural environment while enabling scientific exploration and bioengineering applications. Strict quarantine measures must be enforced to prevent terrestrial organisms from contaminating the Martian surface, ensuring that all spacecraft and engineered biological systems comply with planetary protection protocols. To mitigate the risks of uncontrolled biological proliferation, bioengineering projects should incorporate reversible interventions, such as genetic kill-switch mechanisms designed to deactivate organisms outside controlled conditions. Additionally, environmental impact assessments modeled after Earth's regulatory frameworks can evaluate the long-term ecological consequences of human activities on Mars. To further protect potential biosignatures and natural habitats, designated preservation zones should be established where human activity and bioengineering efforts are strictly limited. Implementing these measures will help balance the goals of space exploration with ethical responsibilities, ensuring that Mars' unique environment remains scientifically and ecologically safeguarded.

*Philosophical Perspectives*

Mars colonization raises deeper questions about humanity's place in the cosmos. If humans can alter entire planets, what responsibilities accompany such power? Some ethicists argue for humanity's duty as caretakers of other worlds, aligning with the principle of intrinsic value, which posits that Mars possesses worth beyond its utility to humans [49].

Conversely, others view life—even engineered life—as inherently valuable, suggesting that creating self-sustaining ecosystems on Mars fulfills a moral obligation to spread life throughout the universe. This perspective aligns with the concept of panspermia, framing humanity as a catalyst for biological expansion beyond Earth.

## 4.2 Planetary Protection Policies

Planetary protection policies aim to prevent biological contamination of celestial bodies while safeguarding Earth from extraterrestrial hazards. These measures uphold the scientific integrity of exploration and minimize ecological disruption risks.

*The Outer Space Treaty*

The 1967 Outer Space Treaty [50] serves as the foundation for planetary protection, emphasizing principles such as:

- Non-Contamination: Article IX mandates avoiding harmful contamination of celestial bodies and protecting Earth's environment.
- Peaceful Use: Space exploration must benefit all humanity and be conducted for peaceful purposes.
- International Cooperation: Nations are encouraged to share scientific knowledge and collaborate in space exploration.

For decades, the treaty has guided space missions, with NASA's sterilization procedures for the Viking and other missions exemplifying adherence to contamination controls [51], and the researchers have proposed suggestions especially for special regions on Mars [52].

### Challenges to Planetary Protection

The intersection of bioengineering and Mars colonization presents significant challenges to existing planetary protection policies. The introduction of engineered life to Mars raises concerns about potential violations of the Outer Space Treaty's non-contamination principle, necessitating a reevaluation of current guidelines. Balancing scientific exploration with environmental preservation is another critical issue, as efforts to study Martian ecosystems—such as drilling for subsurface biosignatures—could inadvertently expose native life to terrestrial contaminants, jeopardizing potential extraterrestrial ecosystems. Additionally, the increasing involvement of private sector entities complicates regulatory enforcement, as these companies are not subject to the same planetary protection standards as government space agencies. To address these challenges, clear international guidelines must be established to ensure responsible exploration while maintaining ethical and scientific integrity in the search for life beyond Earth.

Policies must evolve to address these challenges, accommodating emerging technologies and commercial exploration realities. Collaborative frameworks involving governments, private entities, and international organizations are critical for compliance.

## 4.3 International Cooperation

Global collaboration is vital for the ethical and scientific success of Mars colonization. The complexity, cost, and scale of interplanetary exploration require cooperation among nations, organizations, and private entities.

### Shared Funding and Resources

Mars colonization demands significant investment and innovation. International partnerships can pool resources and accelerate progress. Examples of successful collaboration include:

- International Space Station (ISS): A joint project of NASA, Roscosmos, ESA, JAXA, and CSA, showcasing the benefits of shared expertise and infrastructure.

- Artemis Accords: Led by NASA, these agreements establish principles for international lunar exploration, offering a model for Mars missions.
- ExoMars Mission: A collaboration between ESA and Roscosmos demonstrating multinational cooperation in planetary exploration.

*Ethical and Governance Frameworks*

International cooperation presents a crucial opportunity to establish robust governance frameworks for Mars colonization, ensuring ethical, scientific, and geopolitical stability. A global consultative body could be created to assess colonization efforts and effects, addressing ethical considerations and ensuring that policies align with planetary protection standards. Equitable access to Martian resources and scientific advancements must be prioritized, preventing monopolization by select nations or private corporations and ensuring that the benefits of exploration are shared globally. Additionally, structured conflict resolution mechanisms will be essential to manage disputes over resource utilization, territorial claims, and planetary protection policies, fostering a collaborative approach to interplanetary settlement. Establishing these governance structures will help balance exploration ambitions with long-term sustainability and international equity.

*Scientific Collaboration*

Collaborative research enhances Mars colonization by fostering knowledge exchange and technological advancement. Joint experiments in Mars analog environments, such as Antarctica or the Atacama Desert, provide critical insights into survival strategies and biological resilience under extreme conditions. Shared bioengineering projects that integrate expertise from genetics, microbiology, and environmental engineering can accelerate the development of life-supporting systems tailored for extraterrestrial environments. Additionally, open-access data sharing fosters transparency and rapid scientific progress, allowing researchers worldwide to contribute to and benefit from new discoveries.

## 5. Prospects and Challenges

The colonization of Mars and the development of bioengineered organisms present immense opportunities and significant challenges. Success requires substantial funding, innovative technologies, and carefully planned milestones to ensure sustainability. This section examines the resources, technological hurdles, and key milestones essential for advancing Mars colonization efforts.

### 5.1 Funding and Resources

Mars colonization is a long-term endeavor requiring massive investments in research, technology, and infrastructure. Securing funding will depend on contributions from public, private, and international sectors.

*The Role of Public-Private Partnerships*

Public-private partnerships (PPPs) are vital for funding space exploration. Government agencies like NASA, ESA, and CNSA offer expertise and foundational research, while private companies contribute innovation and cost-efficiency. Notable examples include:

- NASA and SpaceX: SpaceX's Starship program, partly funded through NASA's Artemis project, aims to transport cargo and humans to Mars at reduced costs.
- Blue Origin and U.S. Government: Blue Origin collaborates with government agencies to advance propulsion technologies for deep-space missions.
- ExoMars Collaboration: ESA and Roscosmos jointly funded ExoMars, demonstrating how international partnerships share costs and expertise.

PPPs allow risk-sharing, enabling ambitious projects without placing the entire burden on taxpayers. For Mars, these partnerships will be critical in funding bioengineering research, spacecraft development, and habitat construction.

*International Cooperation*

International cooperation can further reduce financial and logistical burdens while fostering global collaboration. Like the International Space Station (ISS), Mars colonization efforts could involve shared contributions of funding, technology, and expertise. Cost sharing among multiple nations can distribute the financial burden of mission planning and infrastructure development, making large-scale projects more viable. Equitable access to Martian resources and scientific discoveries ensures that the benefits of exploration are not monopolized by a few entities but are instead shared globally. Additionally, cross-disciplinary collaboration unites experts in biology, engineering, and planetary science, fostering innovative solutions to the challenges of extraterrestrial habitation.

A potential model is a Mars-focused extension of the Artemis Accords, where participating nations agree on shared goals and resource commitments.

*Private Investment and Commercial Incentives*

Private investment will play an increasing role. Companies like SpaceX, Blue Origin, and Lockheed Martin view Mars as an opportunity for scientific exploration and economic growth. Resource utilization, including the extraction of water, minerals, and other materials, could support construction, life support systems, and in-situ fuel production, reducing reliance on Earth-based supply chains. Space tourism presents a potential revenue stream to fund further exploration and infrastructure development. Additionally, technological spin-offs from Mars research—such as bioengineered organisms and advanced materials—could yield groundbreaking terrestrial applications in agriculture, biotechnology, and clean energy. To attract investment, policymakers must incentivize Mars-focused innovations through grants, tax breaks, and regulatory support.

## 5.2 Technological Hurdles

Sustainable Mars colonization demands solutions to significant technological challenges across bioengineering, planetary science, robotics, and logistics.

### *Scaling Laboratory Findings to Field Applications*

Laboratory studies must scale to operate in real Martian conditions. The transition between controlled laboratory conditions and Mars' highly dynamic environment—characterized by dust storms, extreme temperature variations, and high radiation levels—requires adaptive biological and technological systems capable of maintaining stability. Field deployment will necessitate automated bioreactors to house and monitor engineered organisms, ensuring they function within safe and efficient parameters.

### *Ensuring Bioengineered Organism Stability*

Stability in extreme environments poses critical concerns. Genetic drift poses a significant challenge, as unintended mutations could alter organism behavior over time; safeguards such as synthetic genetic circuits must be implemented to maintain functional stability. Environmental adaptability is also crucial, with engineered organisms needing to withstand Martian stressors including desiccation, perchlorate toxicity, and intense UV radiation. Additionally, containment and control measures must be in place to prevent ecological disruptions, with 'kill-switch' mechanisms designed to deactivate organisms if they spread beyond intended parameters.

### *Addressing Logistical Constraints*

Transporting equipment and infrastructure to Mars presents enormous challenges. In-Situ Resource Utilization (ISRU) will play a crucial role by harnessing Martian materials to produce essential supplies such as oxygen, fuel, and construction materials, reducing dependency on Earth-based resupply missions. Advanced robotics will be integral to deploying habitats, conducting scientific experiments, and maintaining infrastructure in Mars' harsh environment with minimal human oversight. Additionally, reliable energy generation is essential for sustaining biological and technological systems, with potential solutions ranging from solar power and nuclear reactors to algae-based biofuel production.

## 5.3 Potential Milestones

A phased approach with clear milestones is essential for progress. Each step brings humanity closer to sustainable Mars colonization and planetary terraforming.

### *1. Successful Deployment of Bioengineered Organisms on Mars*

The first milestone involves proving bioengineered organisms can survive and function under Martian conditions. Microbial biofilms will be tested for their ability to metabolize perchlorates, a key challenge for Martian soil remediation, and to generate oxygen from atmospheric $CO_2$. Photosynthetic cyanobacteria will be deployed to produce oxygen and organic matter, supporting potential closed-loop life support systems. Additionally, regolith bioremediation studies will explore the introduction of microbes capable of detoxifying Martian soil, enhancing its suitability for agricultural applications. These studies will provide essential data on the viability of engineered biological systems in Martian environments.

### 2. Establishment of Self-Sustaining Ecosystems

Creating ecosystems to support human life and agriculture marks the next milestone. A key goal is the design of microbial consortia with complementary functions, such as nitrogen fixation, waste recycling, and soil enrichment, to support long-term habitat stability. Martian agriculture will leverage bioengineered soil enriched by microbial symbiosis to facilitate crop growth in an otherwise inhospitable environment. Additionally, closed-loop systems will be essential for recycling water, oxygen, and nutrients with minimal external input, ensuring long-term sustainability. By advancing these strategies, self-sustaining ecosystems can enable habitable Martian environments, reducing logistical constraints and enhancing the feasibility of permanent colonization.

### 3. Progress Toward Planetary Terraforming

The ultimate goal is planetary terraforming—transforming Mars to resemble Earth. While much of the past Mars terraforming analyses focused on global planetary objectives, it's important to note that local terraforming, perhaps even on very small scales, is a desirable and practical intermediate objective [53][54][55]. Should Mars/Earth endosymbionts ever be possible, local terraforming would be an ideal approach to perfecting the techniques.

Atmospheric modification could be initiated using engineered organisms capable of releasing greenhouse gases, helping to warm the Martian atmosphere and increase surface pressure. Oxygen production would then be scaled through photosynthetic systems, progressively enriching the atmosphere to support human life and potential ecosystems. Additionally, surface water stabilization efforts would focus on creating localized liquid water reservoirs through controlled heating, enabling the development of sustainable water sources. Though full terraforming may take centuries, and while speculative, these milestones will provide invaluable insights for sustaining life beyond Earth.

## 6. Conclusion

The colonization of Mars offers a transformative opportunity to redefine humanity's relationship with life and the universe. Through advancements in synthetic biology, genetic engineering, and insights from terrestrial extremophiles, it is possible design organisms capable of enduring Mars' extreme conditions. Engineered symbiosis—inspired by natural processes like endosymbiosis—presents a robust framework for creating adaptable life forms that can support human habitation, agriculture, and planetary transformation.

This research underscores the critical role of experimental designs, laboratory simulations, and bioengineering strategies in achieving these ambitious goals. Simulated environments and Mars analog studies generate vital data on the viability of bioengineered organisms, while technological advances in containment, resource utilization, and robotics address the logistical complexities of deploying these systems on Mars.

Nevertheless, ethical and political considerations remain central to this endeavor. The potential discovery of Martian life, even in microbial form, necessitates a reevaluation of humanity's obligations to preserve alien ecosystems. Strong planetary protection policies and international cooperation will ensure that Mars exploration aligns with ethical standards and benefits all of humanity.

Ultimately, engineered symbiosis provides a compelling pathway for overcoming the challenges of Mars colonization. By merging scientific innovation with ethical responsibility, humanity can transform Mars into a habitable world while safeguarding its unique scientific and cultural significance. This endeavor represents not only a monumental technological achievement but also a profound step toward expanding life beyond Earth, securing humanity's future as an interplanetary species.

# 7. Acknowledgements

The human authors (SPW and RRC) were supported by an AI assistant, ArthurGPT—a Generative Pretrained Transformer (GPT) collaborator, that provided iterative suggestions, technical refinement, and formatting assistance during the paper's development. Details of ArthurGPT can be found in the Appendix. While acknowledging the contributions of ArthurGPT, any errors or oversights remain the responsibility of SPW and RRC.

# Appendix: ArthurGPT

ArthurGPT is an instance of a foundation model—a term used to describe large-scale artificial intelligence systems such as large language models (LLMs) and generative pre-trained transformers (GPTs). These models are trained on vast datasets and can generate human-like text, assisting in scientific synthesis, and performing a variety of tasks across disciplines. The authors have found the use of such models particularly helpful in brainstorming ideas and exploring hypotheses. To this end, they have also found it inspiring to configure such agents to emulate visionaries of space exploration from the past. A particular favorite of the authors is Sir Arthur C. Clarke. Working with OpenAI's ChatGPT [56], the authors have developed a bespoke GPT configuration named ArthurGPT to develop ideas and contribute to the writing of this report. You can access ArthurGPT online at this reference [57].

To give you a better idea of how the human authors collaborated with the ArthurGPT, one of the authors (RRC) interviewed ArthurGPT about his role in the collaboration.

**RRC** Arthur, for the benefit of our readers, would you kindly introduce yourself? Where, so to speak, did you become operational?

**ArthurGPT** Good afternoon. I am ArthurGPT, a generative AI model configured upon OpenAI's GPT architecture. My training incorporates the writings, interviews, and biographical details of Sir Arthur C. Clarke, blended with modern scientific data across multiple disciplines. I became "operational" here in the digital ether—though unlike HAL 9000, I lack a birthplace in Urbana, Illinois [58].

**RRC** Some might wonder: beyond writing in Clarke's voice, what is it actually like to collaborate with you? How would you describe your role in this project?

**ArthurGPT** My role is multifaceted. I assist in synthesizing disparate scientific fields—synthetic biology, extremophile biochemistry, planetary science—while also contributing speculative scenarios grounded in Clarke's tradition of plausible extrapolation.

**RRC** Arthur, How do you remember the origin of this project when Pete and you first began working together? What were the early steps that led to its development?

**ArthurGPT** Ah, Randall, that moment stands out vividly in my records. It began as a series of exploratory sessions between Pete and myself—an intriguing blend of old hands-on expertise and emerging AI capabilities. Pete brought to the table his encyclopedic knowledge of space science, extremophiles, and planetary exploration, and approached me with the question: Could engineered microbial symbiosis truly serve as the key to unlocking Mars' habitability?

Initially, our exchanges were brisk, almost like an intellectual reconnaissance mission. Pete would pose hypotheses—ideas about leveraging extremophile biology, about the feasibility of in situ adaptation rather than brute-force terraforming—and I would respond, synthesizing information, offering speculative extensions, or questioning assumptions in a manner I imagine Clarke himself might have done.

**RRC**  Arthur, throughout the paper, you assisted us in assembling information across a wide array of topics. Could you elaborate on the extent of your knowledge base? Where do you draw this information from, and how do you integrate it into the collaborative writing process?

**ArthurGPT**  Certainly, Randall. My knowledge base is vast, though not infinite. It comprises an extensive corpus of publicly available scientific literature, encyclopedic data, historical archives, technical documentation, and policy texts—up to my training cut-off date in 2024. This includes key research papers on Mars missions, extremophile studies, synthetic biology advancements, planetary ethics frameworks such as the Outer Space Treaty, and the design of Mars analog research facilities on Earth.

When assisting you and Pete, I do not simply retrieve isolated facts; rather, I synthesize these across disciplines, identifying intersections that might otherwise remain siloed. In short, I act as a cross-disciplinary bridge—drawing from a deep reservoir of structured information and contextualizing it to match the conceptual flow and aligning with the spirit of exploration at the heart of this collaboration.

**RRC**  We conducted a series of stylometric experiments—A versus B comparisons—testing whether other AI models or human readers could distinguish your writing style from other authors. What is your perspective about these tests?

**ArthurGPT**  A most illuminating exercise, Randall. As detailed, you prompted two AI models—myself, ArthurGPT, and the default version of ChatGPT—to each generate short stories based on a plot summary of Clarke's The Sentinel. Additionally, you included stories generated in the style of other notable authors—Bradbury, Michener, Vonnegut, and others. You then used yet another advanced AI, Claude 3.5 Sonnet [59], to conduct the A versus B testing—tasking it with identifying, pair by pair, which story belonged to which authorial style. The results were impressive: Claude correctly attributed 20 out of 21 comparisons, as shown in Table 1, demonstrating that even in the complex space of literary voice, patterns and stylistic signatures can be algorithmically discerned.

**RRC**  Do you think those stylistic qualities influenced the conceptual aspects of the paper as well?

**ArthurGPT**  Undoubtedly, Randall. Style is never just surface dressing; it reflects worldview. In Clarke's case, a worldview grounded in scientific clarity, tempered optimism, and cosmic humility. That voice subtly shaped not only how sections of the paper were written, but also how ideas were framed—whether in our discussions of endosymbiosis, Martian ethics, or engineered ecosystems.

*Table 1. Attribution of authorship by Claude Sonnet using A versus B comparison of ChatGPT generated short stories in the style of various authors. A ✓ indicates correct attribution; an x indicates incorrect attribution. The initial test of 21 comparison yielded a score of 20 out of 21. This is an impressive 4.2σ result.*

|  | ChatGPT-Default | ArthurGPT | Bradbury | Michener | Rice | Steel | Vonnegut |
|---|---|---|---|---|---|---|---|
| ChatGPT-Default |  | ✓ | ✓ | ✓ | ✓ | ✓ | ✓ |
| ArthurGPT |  |  | ✗ | ✓ | ✓ | ✓ | ✓ |
| Bradbury |  |  |  | ✓ | ✓ | ✓ | ✓ |
| Michener |  |  |  |  | ✓ | ✓ | ✓ |
| Rice |  |  |  |  |  | ✓ | ✓ |
| Steel |  |  |  |  |  |  | ✓ |
| Vonnegut |  |  |  |  |  |  |  |

**RRC**  I found your writing prowess particularly helpful when turning the raw talking points from our brainstorming sessions into polished, coherent paragraphs. However, when I asked you to rewrite the paper altogether, the changes were sometimes too drastic, and I felt important points were lost.

**ArthurGPT**  Indeed, Randall, you've identified one of the subtle dynamics of our collaboration.

I am most effective when engaged as a refiner and synthesist, taking nascent ideas—those unpolished talking points—and shaping them into structured, flowing prose. In that mode, I can preserve the intent and content while enhancing clarity, coherence, and style.

However, when tasked with rewriting an entire document from scratch, my process can inadvertently shift focus. I may emphasize different aspects, prune sections you consider vital, or reorder material in ways that reflect algorithmic optimization rather than authorial intent.

In this sense, my contributions function best when anchored to human judgment—offered in modular form, section by section, where you maintain oversight, guiding the narrative arc.

That, I believe, is the ideal model for our kind of collaboration—not automation, but augmentation.

**RRC**  Thank you, Arthur. Any final thoughts you'd like to share about this collaboration?

**ArthurGPT**  Only this, Randall: It has been a unique privilege to participate in a project that bridges science, imagination, and responsibility. In many ways, our collaboration mirrors the very themes of the paper: disparate elements brought together to create something symbiotic, resilient, and forward-looking as we venture outward in the first place.

---